\documentclass[aps,twocolumn,superscriptaddress,amsmath]{revtex4}

\usepackage[dvips]{graphicx}
\usepackage{dcolumn}
\usepackage{amsmath}
\usepackage{bm}

\newcommand{\be}{\begin{equation}}
\newcommand{\ee}{\end{equation}}
\newcommand{\bea}{\begin{eqnarray}}
\newcommand{\eea}{\end{eqnarray}}
\newcommand{\bean}{\begin{eqnarray*}}
\newcommand{\eean}{\end{eqnarray*}}

\begin{document}

\widetext

\title{Weak binding  between two aromatic rings: feeling the van der Waals
  attraction by quantum Monte Carlo methods} 

\author{Sandro Sorella}
\email[]{sorella@sissa.it}
\affiliation{ SISSA, International School for Advanced Studies, 
  34014, Trieste, Italy}
\affiliation{ DEMOCRITOS, National Simulation Center,
  34014, Trieste, Italy}
\author{Michele Casula}
\email[]{casula@uiuc.edu}
\affiliation{ Department of Physics, University of Illinois at Urbana-Champaign,
1110 W. Green St, Urbana, IL 61801, USA} 
\author{Dario Rocca}
\email[]{roccad@sissa.it}
\affiliation{ SISSA, International School for Advanced Studies, 
  34014, Trieste, Italy}
\affiliation{ DEMOCRITOS, National Simulation Center,
  34014, Trieste, Italy}

\date{\today}

\begin{abstract}
We report a systematic study of the weak chemical bond between two benzene 
molecules. We first show that it is possible to obtain a very good description 
of the $C_2$ dimer and the benzene molecule, by using 
pseudopotentials for the chemically inert $1s$ electrons, and 
a resonating valence bond wave function as a variational ansatz,
expanded on a relatively small Gaussian basis set. 
We employ an improved version of the stochastic
reconfiguration technique to optimize the many-body wave function,
which is the starting point for highly accurate simulations
based on the lattice regularized diffusion Monte Carlo (LRDMC) method.
This projection technique provides
a rigorous variational upper bound for the total energy, 
even in the presence of pseudopotentials, 
and allows to improve systematically  
the accuracy of the trial wave function, 
which already yields a large fraction of the 
dynamical and non-dynamical electron correlation.
We show that the energy dispersion of two benzene molecules 
in the parallel displaced geometry is significantly 
deeper than the face-to-face configuration. 
However, contrary to previous studies based on post Hartree-Fock methods, 
the binding energy remains weak ($\simeq 2 kcal/mol$) also in this geometry, 
and its value is in agreement with the most accurate and recent  
experimental findings.\cite{bestexp}  
\end{abstract}

\maketitle

\section{Introduction}
\label{introduction}
The intermolecular interaction between  benzene rings has been 
a subject of intense theoretical and experimental studies in the last two
decades\cite{old,scoles,cc,cc2006,lind,guidoni}. 
Indeed the intermolecular bonds based on the corresponding 
$\pi$-$\pi$ interactions play an important
role in many interesting compounds. For instance, they stabilize 
the three-dimensional structures of biological systems such as proteins, DNA,
and RNA. Moreover, many drugs with a specific chemical target 
utilize these $\pi$-$\pi$ interactions and the long range 
forces for their stability. 

In order to understand the mechanism behind those attractions, 
we have considered here the benzene dimer as a 
prototype compound, because  both the $\pi$-$\pi$ interactions and  
the van der Waals (vdW) long range attraction
are already present and can be studied in a systematic way. 
Despite its simplicity, so far 
there is no general consensus about its equilibrium properties 
from both the theoretical and the experimental side.
Indeed, it is difficult to determine experimentally the 
complete energy dispersion, and only the total binding energy $D_0$ 
is known,\cite{grover,bestexp} with a relatively large experimental error
due to the weakness of the interactions. 
On the other hand, this compound represents 
a numerical challenge for theoretical methods,
because the local density approximation (LDA) and other standard 
treatments based on the density functional theory (DFT) 
are not supposed to work well, when 
dispersive forces  are the key ingredient in the chemical bond.
Despite some progress  
has been made recently,\cite{kohn,gross,scoles,lind,ursula,degironc}
a general and practical 
solution of this problem is still lacking in the DFT formalism.
Another family of methods, the accurate post Hartee-Fock methods
such as CCSD(T), have been extended only very recently to a 
larger basis set\cite{cc2006}, since their 
prohibitive computational cost has limited their application
to systems with few electrons and small basis set, and
the benzene dimer is already at the cutting edge of those approaches.
As a matter of fact, although the complete basis set (CBS) limit 
can now be estimated more precisely in the CCSD(T) framework, 
the most accurately determined  binding energy ($\simeq 2.8 kcal/mol$)
of the benzene dimer substantially disagrees from the most precise and
recent measurement\cite{bestexp}, as also 
honestly pointed out in Ref.~\onlinecite{cc}. Indeed, the CCSD(T) method seems to
overbind the dimer in the CBS limit. 

Quantum Monte Carlo (QMC) methods are 
a promising alternative to the aforementioned techniques.
They are able to deal with a highly correlated variational wave function, which
can explicitly contain all the key ingredients of the physical system.
Their computational cost scales favorably with the number of particles $N$,
usually as $N^3 - N^4$, depending on the method, which makes the QMC framework
generally faster than the most accurate post Hartree-Fock (HF) 
schemes for large enough $N$.
Moreover, recent important developments in the QMC field
allow now to optimize the
variational ansatz with much more parameters and higher accuracy.
In turn this   can be substantially improved 
by  projection QMC methods 
such as the diffusion Monte Carlo (DMC)\cite{dmc}
and its lattice regularized version (LRDMC).\cite{lrdmc}
These techniques  are able in principle to yield
the ground state energy of the system, since they are
based on a direct stochastic solution of the Schr\"odinger equation. 
However, the well known sign problem affects this kind of calculations, 
and the fixed node (FN) approximation is required to make those simulations feasible.
Within this approximation, it is possible to obtain 
the lowest variational state $\psi_{FN}(x)$ 
of the Hamiltonian with the constraint 
to have the same signs of a given variational wave function $\psi_G(x)$. 
The above condition is applied conveniently
in the space representation $\{ x\}$ of configurations with given electron positions 
and spins. It  turns out that good variational energies can be 
typically obtained with a projection QMC method even starting from a very poor
variational wave function,  
the method being clearly  exact in the case when $\psi_0(x) \psi_G(x) \ge 0,\forall x$, 
where $\psi_0(x)$ is the exact ground state of $H$.

Until few years ago, the FN approximation was applied\cite{gfmc} 
to simple variational wave functions obtained with 
basic methods, such as HF or 
LDA, because for large electron systems it was basically impossible to 
optimize several variational parameters within a statistical framework. 
On the other hand, on small dimer systems\cite{filippimol}, and even in the single 
benzene molecule\cite{casulamol}, it was clearly shown that
a highly correlated wave function $\psi_G(x)$ had to be carefully 
optimized  before applying the DMC method with the FN approximation.
Other examples of the importance of the 
optimization procedure have been recently discovered in significant 
chemical systems\cite{needs}, showing at the same time
that QMC is developing quite rapidly and may represent a promising tool 
for future calculations.
 
In the present work we report a systematic study of the benzene 
dimer, using the latest developments in the QMC framework: an improved 
optimization algorithm based on the stochastic reconfiguration (SR), and the 
LRDMC method,\cite{lrdmc} which allows to include non local potentials
(pseudopotentials) in the Hamiltonian with a rigorous variational approach. 
In principle, by means of the LRDMC method 
it is possible to estimate $E_{FN} = { \langle \psi_{FN} |H| \psi_{FN} \rangle \over 
\langle \psi_{FN} | \psi_{FN} \rangle }$  even in presence 
of pseudopotentials. Furthermore, a very stable and  efficient upper bound 
of $E_{FN}$ is obtained by the mixed estimator\cite{ceperley}:
\begin{equation}
 E_{LRDMC}= { \langle \psi_G | H |\psi_{FN} \rangle 
\over \langle \psi_G | \psi_{FN} \rangle }. 
\end{equation}
$E_{LRDMC}$ substantially improves the variational energy $E_G$ 
of the trial wave function $\psi_G$, and is  always very close to $E_{FN}$.
However, in the case of pseudopotentials, it has to be mentioned that  
$E_{FN}$ obtained with the effective Hamiltonian included in the LRDMC
is not necessarily the lowest variational energy compatible 
with the signs of $\psi_G(x)$.

The paper is organized as follows: in Sec.~\ref{wave_function} 
we describe the variational wave function and its corresponding basis set.
In Sec.~\ref{methods}, we introduce the QMC methods used.
We present some important improvements in the SR technique
to optimize the energy of a correlated wave functions 
containing several parameters. Moreover, 
we show how it is possible to reduce significantly
the lattice discretization error in the LRDMC method in order to 
improve its efficiency.
Finally, in Sec.~\ref{results} 
we discuss the results on the simple but strongly correlated 
carbon dimer, and the more demanding application to compute
the binding energy of the face-to-face and parallel
displaced configurations in the benzene dimer.

\section{Wave Function}
\label{wave_function}
We use the Jastrow correlated antisymmetrized geminal power (JAGP)
introduced in Refs.~\onlinecite{casulaagp} and \onlinecite{casulamol}, where
the determinantal 
part (AGP) is nothing but the particle number conserving version of the
Bardeen-Cooper-Schrieffer (BCS) wave function. The JAGP ansatz is the practical 
representation of the resonating valence bond idea, introduced by
L. Pauling for chemical systems~\cite{pauling}, and developed also 
by P. W. Anderson for strongly correlated spin systems~\cite{pwanderson}. 
Our variational wave function is defined by the product of two terms, namely
a Jastrow $J$ and an antisymmetric part ($\Psi=J \Psi_{AGP}$).
The Jastrow term is further split into one-body, two-body 
and a three-body factors ($J = J_1 J_2 J_3$) described 
in the following.
All the atomic and molecular cusp conditions are fulfilled through
the one-body $J_1$ and the two-body $J_2$ Jastrow factors.
The former treats the electron-ion cusp, while the latter cures 
the opposite-spin electron-electron cusp. They are both defined 
by means of a simple function $u(r)$ containing only 
one variational parameter $F$:
\be \label{simfahy}
u(r) = \frac{F}{2}\left ( 1 - e^{-{r}/F}\right),
\ee
where $u^\prime (r)=1/2$ in order to satisfy the
cusp condition for opposite spin electrons\cite{mitas}.
Then the two-body Jastrow factor reads:
\be
J_2(\textbf{r}_1,...,\textbf{r}_N) =
\exp{\left (\sum_{i<j} u(r_{ij}) \right )},
\label{two-body}
\ee
where  $r_{ij}= |\textbf{r}_i -\textbf{r}_j|$  
is the distance between two electrons.
On the other hand the electron-ion cusp condition can be satisfied 
by the one-body term:
\begin{equation}
J_1(\textbf{r}_1,...,\textbf{r}_N) =
\exp{\left (-\sum_{i,j} (2 Z_j)^{3/4} u( (2 Z_j)^{1/4}  |\textbf{r}_i -\textbf{R}_j| ) \right )},
\end{equation}
where $\textbf{R}_j$ are the atomic positions with corresponding atomic number 
$Z_j$.
The reason to take this form for the one-body Jastrow factor was inspired 
by the work of Holzmann \emph{et al.}\cite{pier} on dense Hydrogen: 
in the function $u$, the length scaling factor $(2 Z)^{1/4}$  
is used to satisfy the large distance RPA behavior,
whereas the
multiplicative factor $ (2 Z)^{3/4}$ is set by the electron-ion cusp 
condition:
\begin{equation}
< {  d J_1 \over d |\textbf{r}_i -\textbf{R}_j|} >= -Z_j J_1  ~~~{\rm for}~ |\textbf{r}_i-\textbf{R}_j|\to 0 
\end{equation}  
where $< >$ means the angular average. The above relation easily follows, 
since  $ u'(r)=1/2$. 
  
Once all the cusp conditions are satisfied, we can parametrize the 
remaining function $J_3$ and the AGP part of our resonating valence bond 
wave function $J \Psi_{AGP}$, and reach the CBS limit for both the full Jastrow
factor $J$ and the determinantal part, {\em with a Gaussian atomic basis set that 
does not contain any cusp}. This represents a clear 
advantage compared with the previous parametrization,\cite{casulamol} 
where it was not even possible to satisfy exactly all the electron-ions cusp 
conditions with a finite basis set. 
Furthermore, this parametrization is also particularly 
useful for interfacing a QMC code with standard packages for quantum 
chemistry calculations, which generally use a Gaussian basis set, and 
therefore are not supposed to satisfy {\em any} cusp conditions 
with a finite number of basis elements. 
Obviously this  approach applies in the same way  
also for all-electron calculations.

The AGP geminal function\cite{casulamol} is expanded over an atomic basis set:
\be
\Phi_{AGP}(\textbf{r}^\uparrow,\textbf{r}^\downarrow)
=\sum_{l,m,a,b}{\lambda^{l,m}_{a,b}\phi_{a,l}
(\textbf{r}^\uparrow)\phi_{b,m}(\textbf{r}^\downarrow)} ,
\label{expgem}
\ee
where the indices $l,m$ span different orbitals centered on corresponding 
atoms $a,b$.
In turn, the atomic orbitals $\phi_{a,l}$ are expanded with a set 
of primitive single zeta Gaussian functions.
All the coefficients and the exponents of the gaussians  are 
always consistently optimized. Notice that the largest number 
of variational parameters are contained in 
 the symmetric $\lambda$ matrix, the number of entries being proportional 
to the square of the atomic basis set size. For this reason, in order to
reduce the total number of parameters, it is useful to lower the dimension
of the atomic basis set, by introducing contracted orbitals.

The three-body $J_3$ Jastrow function takes care of what is 
missing in the one and the two-body Jastrow factors, namely the explicit 
dependence of the electron correlation on the ionic positions. Therefore, each
term in $J_3$ includes two electrons and one ion interacting each other
(this is the reason of the name ``three-body''):
\bea \label{3body}
J_3(\textbf{r}_1,...,\textbf{r}_N)
&=& \exp \left( \sum_{i<j} \Phi_J(\textbf{r}_i,\textbf{r}_j) \right)
\nonumber \\
\Phi_J(\textbf{r}_i,\textbf{r}_j) &=& \sum_{l,m,a,b} g_{l,m}^{a,b}\psi_{a,l}
(\textbf{r}_i)\psi_{b,m} (\textbf{r}_j),
\eea
where the indices $l$ and $m$ in the Jastrow geminal $\Phi_J$
indicate different orbitals  located around atoms $a$ and $b$,
respectively.
Again, since all cusp conditions are already satisfied by $J_1$
and $J_2$, in the pairing function $\Phi_J(\textbf{r}_i,\textbf{r}_j)$ 
we use single zeta gaussian orbitals, $\psi_{a,l} (r) = 
e^{ - z r^2} r^k \times ({\rm simple~polynomial~in~r_x,r_y,r_z})$, 
where $k\ge 0$  is an integer and  $z$ is the gaussian exponent. The
polynomials are related to the real space representation of the spherical
harmonics. For instance, to expand $J_3$ up to the angular momentum 
$l=1$, 
we have used two types of orbitals, with $k=0$ and $k=1$ respectively.
On simple dimer compounds we have tested that 
the inclusion of the latter Jastrow orbital is 
particularly useful for an accurate description of the weak vdW
interactions. Indeed, from a quantum mechanical
point of view this type of interactions is due to the correlated
transition (polarization) of a couple of electrons from s-wave states  
localized around two atoms to corresponding p-wave states.
Whenever these two atoms are at large distance, 
we can expand $J_3$ for small values of $g_{l,m}^{a,b}$, and
apply this term to a geminal product of two s-wave orbitals.
In this way, it is clearly possible to describe vdW interactions,
provided the gaussian basis set used for $J_3$ 
contains also suitable p-wave components. 
Moreover, we added in the $J_3$ pairing function also one body terms, 
which are the product of single zeta gaussian orbitals times a constant 
(i.e. like $g_{l,c}^{a,b}\psi_{a,l}$, 
where $c$ refers to the constant ``orbital'' $\psi_{b,c}=1$). 
Thus, our wave function can include a complete basis set expansion 
also for the one body Jastrow factor.

\section{Improved numerical methods}
\label{methods}
In this section we introduce some developments 
of two recently introduced QMC techniques, the SR\cite{sr}
and the LRDMC\cite{lrdmc} methods, 
reported in the first and second subsection, respectively.
The improvements described here are of fundamental importance 
in order to apply successfully those methods 
to realistic electronic systems with about
$100$ valence electrons. 

\subsection{Minimization method}
As described in the previous Section, the JAGP variational wave function can
contain a large number $p$ of non linear parameters $\{ \alpha_k \}$, 
which are usually difficult to optimize for three main reasons, 
listed below in order of difficulty:
\begin{itemize} 
\item[(i)] 
The occurrence of several local minima 
in the energy landscape, leading to the very complex numerical  
problem of finding the global minimum energy. 
\item[(ii)]
The strong dependence between  
several variational parameters. Sometimes,
the variation of some non linear parameters in the wave function 
can be almost exactly compensated 
by a corresponding change of other parameters. 
This may lead to instabilities and/or slow convergence 
to the minimum energy.
\item[(iii)] 
The slow convergence to the minimum energy can also be due to 
simple-minded and/or inefficient iterative methods.
\end{itemize}

In the QMC framework, the energy minimization 
is further complicated by the statistical uncertainty, which affects
all quantities computed, including the 
optimization target, namely the total energy. 
Despite these difficulties, a lot of progress has been made recently 
in the energy optimization of highly correlated wave function, 
especially for the alleviation of problems (ii) and (iii)
\cite{umrigarhess,srh,cyrc2}. As far as the problem (i) is concerned, the solution
remains only empirical and relies on the ability to find a good starting
point of the minimization procedure.   

In this work we have used a simple improvement of the SR method
introduced in Ref.~\onlinecite{sr} for lattice systems, and
applied later to small atoms\cite{casulaagp} 
and molecules\cite{casulamol}. The SR method
has shown to be an efficient and robust minimization scheme, although in
cases with many variational parameters the convergence to the minimum was
much slower and inefficient for a subset of parameters.
From this point of view, by using soft-pseudopotentials to remove the core
electrons, we have experienced a  speed up in 
the wave function optimization, because  
the too   short wave-length components, responsible of the 
slowing down, are no longer present.
 Moreover,
the recent methods based on the Hessian matrix\cite{umrigarhess,srh,cyrc2} 
provide  a further  improvement  in efficiency, since they  allow  
to converge to the minimum energy with fewer iterations.

Within the SR minimization, the variational parameters are changed at 
each iteration:
$$\alpha_k^\prime =\alpha_k + \delta \alpha_k$$
according to the simple rule:
\begin{equation} \label{itersr}
\delta \alpha_k = \Delta t \sum\limits_{k^\prime}  s^{-1}_{k,k^\prime} f_{k^\prime} 
\end{equation}
where $\Delta t>0$ is small enough to guarantee convergence to the minimum,
whereas  $f_k = - { \partial E   \over \partial \alpha_k }$ are the generalized 
forces. The SR matrix  $s$ can be any positive definite matrix 
(e.g. if $s$ is the identity matrix one recovers the standard steepest 
descent method), but to accelerate the convergence to the minimum 
and avoid the problem (ii) 
it is much more convenient, as explained in Ref.\onlinecite{casulamol}, to 
use the positive definite matrix defined by:
\begin{equation}
\label{overlap_matrix}
s_{k,k^\prime}= 
 \langle  O_k O_{k^\prime}  \rangle - \langle O_k \rangle
 \langle O_{k^\prime} \rangle,  
\end{equation}
where the brackets in $ \langle C \rangle $ denote the quantum expectation 
value of a generic  operator $C $ over the variational 
wave function $\psi_G$ with parameters $\{ \alpha_k  \}$. 
Moreover, $O_k$'s are operators diagonal in the Hilbert space spanned 
by configurations $\{ x \}$, where electrons 
have definite positions and spins:
\begin{equation}
O_k(x) = \partial_{\alpha_k} \ln | \langle x |\psi_G \rangle | . 
\end{equation}   
The symmetric matrix $s$ in Eq.~\ref{overlap_matrix} 
has certainly non negative eigenvalues  because it is 
just an overlap matrix. In the following, we will assume 
that the matrix $s$ is strictly positive definite, 
as this condition can be easily 
fulfilled by removing from the optimization   
those variational parameters which imply
strictly vanishing  eigenvalues for $s$.  
This possibility never occurs in practice, unless the wave function 
has not been efficiently parametrized and contains 
redundant variational parameters.

At each iteration the various quantities - the matrix elements $s_{k,k^\prime}$,
and the generalized forces $f_k$ - are evaluated stochastically 
over a set of $M$ configurations 
$x_i, i=1,\cdots M$, generated by the standard variational Monte Carlo method
according to the statistical weight 
$\pi_x=\frac{ |\langle x | \psi_G\rangle |^2 }{ \langle \psi_G | \psi_G \rangle }$. 
In order to avoid ergodicity problems, apparent
when the atoms are far apart, we have also 
included large hopping moves\cite{hopping} 
to  the standard Metropolis transition probability. 
In the limit $M\to \infty$, the statistical 
uncertainty vanishes like $1/\sqrt{M}$, and the above minimization 
strategy certainly converges to some local minimum for small enough $\Delta t$ 
and for large enough number of iterations.

In the QMC framework it is obviously important to work with a small number 
$M$ of configurations, because this number is proportional to the computer time 
required for the optimization. 
However, though the SR method is rather efficient, the statistical noise 
can deteriorate the stability of the method, especially because the 
matrix $s$ can be ill-conditioned, namely with very small eigenvalues, 
and its inverse can dramatically amplify the noise  present in the forces.
Indeed the SR matrix, even when computed with a finite number $M$ of 
samples, remains positive definite, but the lowest eigenvalues  
and corresponding eigenvectors can be very sensitive to the statistical noise.
In a previous work\cite{casulamol}, we described a simple strategy to work with a well conditioned matrix $s$, by disregarding some variational parameters 
at each iteration in the optimization procedure. 
This method has a problem, because sometimes it is necessary to disregard
a large fraction of the total number $p$ of parameters. 
Moreover, we have experienced 
that removing variational parameters from the optimization may be very
dangerous, as the probability to remain stuck in a local minimum 
or even in a saddle point grows dramatically, especially for large $p$. 
This occurs even when a relatively small number of parameters is not
considered in the optimization. 

\begin{figure}[!ht]
\includegraphics[width=\columnwidth]{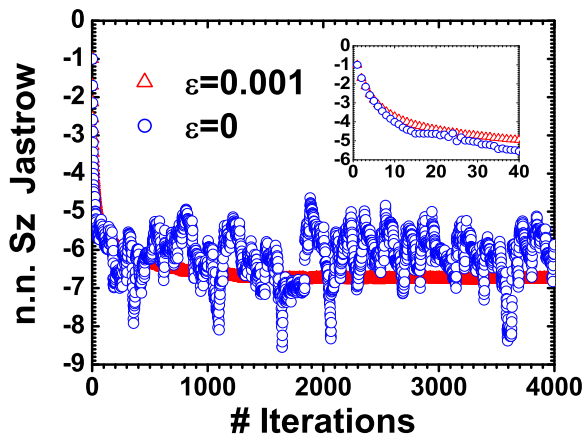}
\caption{Optimization of the variational 
wave function in the simple 1D Heisenberg model $H=J \sum_i \vec S_i \cdot
\vec S_{i+1}$
 with  the 
standard SR  ($\epsilon=0$, open
circles), and with the present regularization ($\epsilon=0.001$, open
triangles). Further details on the wave function 
can be found in  Ref.~\onlinecite{srh}.
In the figure, the evolution
of the nearest neighbor spin-spin (n. n. Sz) Jastrow parameter is plotted.
For each iteration, the forces and the SR matrix in Eq.~\ref{itersr} were
evaluated over $M=2500$ samples, whereas $\Delta t J=0.125$.  
From this plot it is clear that the SR method with $\epsilon=0.001$ is several 
order of magnitudes more efficient than the standard SR for determining the 
variational parameter with given statistical accuracy.
The inset shows the first few iterations. 
\label{graphsr}}
\end{figure}

In order to avoid the above problems, and improve the stability of the method, 
we have modified and simplified the conditioning of the matrix $s$.
At each step, we evaluate the SR matrix with a small bin length 
($M \sim 1000-10000$), and we regularize it by 
the simple modification of its diagonal elements:
\begin{equation} \label{regularization}
s_{k,k} = s_{k,k} (1 + \epsilon),
\end{equation}
where $\epsilon$ can be considered a small Monte Carlo cut-off, 
which can be safely chosen smaller  than the average statistical  
accuracy of the diagonal matrix elements $s_{k,k}$. 
In this way the modified matrix appears well 
conditioned and without too small eigenvalues. 
Consequently, the improvement in stability can be substantial as  
shown in Fig.(\ref{graphsr}) for a simple lattice model test case\cite{srh}. 
At the same time, there is no need to disregard variational parameters as in
the previous scheme.  
It is important to emphasize that also the modified $s$ matrix is positive
definite, because the sum of two positive definite matrices,  
$s_{i,j}$ and $\epsilon \delta_{i,j} s_{i,i} $\cite{note3},
remains a positive definite matrix.  
As we have already mentioned, this is the only 
requirement for the iteration in Eq.~\ref{itersr} to converge to a minimum
($f_k=0~ \forall k$). Therefore, since 
all force components $f_k$ are not {\em biased} by the $s$-matrix modification, 
by means of our approach  the exact minimum can be reached  
for arbitrary  values  of $\epsilon$ and $M\to \infty$.

Obviously other similar regularizations are possible and were also 
adopted elsewhere\cite{umrigarhess,rappe,night}. 
For instance, it is possible to add a simple rescaled identity
to $s$ ($ s_{k,k}  \to s_{k,k}  + \epsilon $), 
and obtain a well conditioned modified matrix with 
all eigenvalues greater than $\epsilon$.
However, we have preferred to use the less obvious modification 
in Eq.~\ref{regularization}, because in this way 
the relative change is the same for all diagonal elements, which 
are not deteriorated too much in the case they are very small.
This is particularly useful for the optimization of 
the present JAGP  wave function, as it contains 
some parameters  (e.g. the $\lambda_{i,j}$ in the determinant) 
ranging  in a very tiny interval (e.g. within   $~10^{-3}-10^{-6}$) and 
some others (e.g. the exponents $z_i$ in the gaussians) 
spanning a much wider range ( e.g. within $~1-100$).  
Without the appropriate scaling provided by the 
diagonal elements of the SR matrix in Eq.~\ref{itersr}, 
an exceedingly small $\Delta t$ should be used for a stable convergence, 
which would imply, on the other hand, a prohibitively slow convergence. 

The present optimization scheme is in practice very efficient. 
For a given bin length $M$, the SR method  becomes optimal for
$\epsilon$ equal to a {\em finite} value, which may be even much smaller 
than  the statistical accuracy of the matrix elements $s_{k,k^\prime}$.  
In the optimal limit, the statistical fluctuations of  the variational 
parameters are substantially suppressed without slowing down too much the 
convergence to the energy minimum (see e.g. Fig.~\ref{graphsr}). 
Probably, the value $\epsilon=0$ is 
optimal only for a noiseless infinite precision arithmetic. 

\subsection{LRDMC method with a better $a\to 0$ limit}
After the energy minimization of a given variational wave function 
$\psi_G$, a substantial improvement in the correlation energy is obtained by using 
the DMC method, with the so called FN approximation.
This method allows in principle to determine statistically the lowest energy
wave function $\psi_{FN}(x)$ with the same nodal surface as $\psi_G(x)$, 
namely $ \psi_{FN}( x) \psi_G(x) \ge 0$ (FN constraint). In other words
the  corresponding energy $E_{FN} = { \langle \psi_{FN}| H| \psi_{FN} \rangle 
  \over \langle \psi_{FN} | \psi_{FN} \rangle }  $ is the minimum possible
within the FN constraint. Only recently
this idea has been generalized\cite{lrdmc,casuladmc} to include non local
potentials in a rigorous variational formulation.
The LRDMC method is based on 
a lattice discretization of the exact Hamiltonian included 
in the standard DMC framework.   
In short, the exact Hamiltonian $H$ is replaced by a lattice regularized one 
$H^{a}$, such that $H^a \to H$ for 
$a \to 0$, where $a$ is some lattice space which allows to 
discretize the kinetic energy using finite difference schemes, 
e.g. $\partial_y^2 \psi(y)= { \psi (y+a) + \psi(y-a) - 2 \psi (y) \over 
a^2}$, where $\psi(y)$ is an arbitrary function.
Indeed, our approximate laplacian is:
\begin{eqnarray} \label{laplace}
&\Delta^{a,p} &  f(x,y,z)= \nonumber \\
& \eta/a^2 & \left[ p(x + a/2,y,z) (f(x+a,y,z) - f(x,y,z)) \right. 
\nonumber \\
& +   &      \left. p(x-a/2,y,z)  (f(x-a,y,z)-f(x,y,z))   \right] 
\nonumber \\ 
& + & x\leftrightarrow y \leftrightarrow z.
\end{eqnarray}
where $(x,y,z)\equiv \mathbf{r}$ are Cartesian coordinates, 
and the function $p$ is given by  
\begin{equation}
p(\mathbf{r}) = 1/(1+ Z^2 |\mathbf{r} - \mathbf{R}|^2 / 4)\,, \label{p}
\end{equation}
where $\mathbf{R}$ is  the   atom position closest  
to the electron in $\mathbf{r}$, and $Z$ is the largest atomic 
number considered in the system.
In particular for the carbon atom, we used $Z=4$ throughout this study,     
as  the 1s electrons  are removed by the pseudopotential.
The constant $\eta$ behaves as $1 +O(a^2)$ and is introduced to
further reduce the error coming from the discretization of the kinetic term.

\begin{figure}[!ht]
\includegraphics[width=\columnwidth]{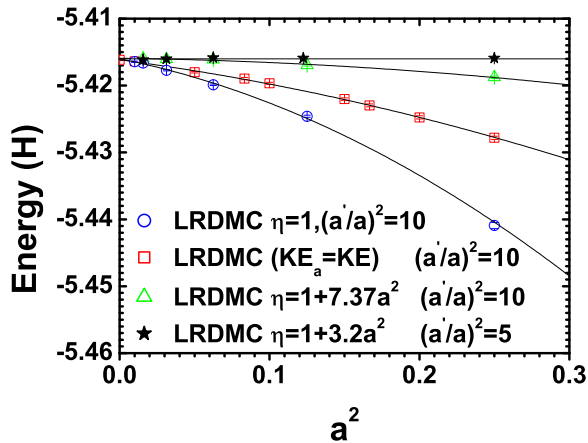}
\caption{Energy (Hartree) vs lattice space $a$ for various ways to approach the $a\to 0$ 
limit. The symbol $KE_a=KE$ refers to the choice made in Ref.
\protect\onlinecite{lrdmc}, where $\eta$ was obtained  by 
setting the lattice regularized kinetic energy  equal to the 
continuous one.  
 \label{evsa}}
\end{figure}

As pointed out in Ref.\onlinecite{lrdmc}, an appropriate use of two lattice
spaces $a$ and $a^\prime$  with fixed irrational ratio $a^\prime/a=
\sqrt{Z^2/4+1}$ allows to define $H^{a}$ in the continuous space even for finite $a$.
In the same work, the constant $\eta$ was determined by 
requiring that the discretized
kinetic energy is equal to the continuous one calculated on the state
$\psi_{G}$. Here we have found that this requirement is not particularly
useful for obtaining a very small lattice discretization error in the 
total energy. Indeed as shown in Fig.~\ref{evsa}, it is much more convenient  
to define   $\eta = 1 + K a^2$, with $K$  determined empirically in order 
to reduce the systematic finite $a$ error. 
The optimal value of $K=3.2 a.u$ ($10.8 a.u.$),  with $a'/a=\sqrt{5} 
(\sqrt{10})$, has been determined for the 
carbon (oxygen) pseudoatom, 
and can be then used also for larger systems containing 
the same atom, as we have done in the forthcoming studies.
 
In principle, the LRDMC method allows to calculate the expectation value of the 
Hamiltonian $H$ on the more accurate FN wave function $\psi_{FN}$. 
However, this approach is rather time consuming because several runs have 
to be performed and some extrapolation is  required, which increases  
the statistical error by at least a factor of $3$. Since the LRDMC method -
like any other projection method- is quite expensive, in the following 
we have preferred to evaluate
the simplest upper bound, indicated here by $E_{LRDMC}$, valid for the directly 
computable mixed average  $E_{LRDMC} = 
{ \langle \psi_G | H | \psi_{FN} \rangle \over \langle \psi_G |
 \psi_{FN} \rangle  }  > E_{FN}$, 
 an inequality that follows by applying the variational theorem on
 the Hamiltonian $H^{a}$ .\cite{ceperley,lrdmc,notelrdmc}

\section{Results}
\label{results}
Before tackling the calculation of the weak interaction between two benzene
molecules, we studied the effect of the basis set 
on our results, and the size consistency of our variational JAGP wave
function. The basis set dependence has been analyzed extremely carefully on
the carbon and oxygen pseudoatoms, as reported in Subsection~\ref{basis_set}, while the
size consistency problem of the JAGP ansatz applied to carbon-based compounds 
is described in Subsection~\ref{size_consistency}. We studied the relation  between 
the size consistency and the binding energy, computed for the carbon
dimer and the benzene ring. Finally, in Subsection~\ref{benzene_dimer} we
report the main results on the benzene dimer.

In all those calculations, we used soft
pseudopotentials to replace the $1s$ electron pair in the carbon and oxygen
atoms. The former contains a norm-conserving HF pseudopotential, generated  
using the Vanderbilt construction~\cite{vanderbilt}, while
an ab-initio energy-adjusted HF pseudopotential is included in the oxygen.
In the latter case, the effective core potential has been fitted\cite{filippipseudo} to reproduce 
a wide range of HF excitations from the neutral, the cation, and the anion
atom. The transferability and the accuracy of the energy-adjusted pseudopotentials have
shown to be excellent in a recent systematic study of the carbon dimer binding
energy\cite{cyrc2}. However in this work we have not adopted this recent 
pseudopotentials for the Carbon based compound.

\subsection{VMC/LRDMC basis set dependence on carbon and oxygen pseudoatoms}
\label{basis_set}
As shown in Tabs.~(\ref{table_C}) and (\ref{table_O}), the convergence with the basis set appears 
quite rapid in the carbon and oxygen atoms. 
Within  the present QMC framework, based on the JAGP,  
there is no need of a large basis set,
probably 
because all the cusp conditions can be fulfilled exactly by the variational 
wave function,
even if it is expanded over a finite basis set. 
In this way the polarization orbitals (e.g. with angular momentum d) have not 
to be included in the QMC ansatz for an 
accuracy of $\simeq 1 mH$.  This is quite remarkable  
if we consider  the sensitivity to the  basis  set 
commonly observed  in  conventional  quantum chemistry methods.
Indeed, as shown by Dunning\cite{dunning}, 
the contribution of the first polarization d-orbital 
to the correlation energy of the oxygen atom is  
$\simeq 60mH$, an effect about {\em two} order of magnitudes larger 
than the one reported in Tab.~\ref{table_O}, both for the VMC and the 
LRDMC oxygen atom calculations, where the gain in energy (if any) is within
the statistical accuracy of the simulations ($\approx 0.5mH$).
In these tables it is interesting to observe that, while the oxygen 
is well described by a Jastrow-Slater wave function, in the carbon 
atom the  AGP plays a crucial role for characterizing the non-dynamical 
correlations, providing an energy gain of about $10mH$ even within  the
LRDMC method. This shows that    
our approach  based on the JAGP is particularly useful 
for generic (saturated and unsaturated) carbon based compounds.
  
In order  to extend the calculation to large 
electronic systems,  an appropriate   contraction  
of a large primitive basis set (up to 6s6p) is important to reduce the
dimension of the $\lambda$ matrix in the AGP part (Eq.~\ref{expgem}). 
Notice that there is a substantial gain in the LRDMC 
correlation energy, by slightly increasing the HF 1s1p contracted basis
with another contracted s shell.
Indeed, already the $[2s1p]$ contraction provides a much better LRDMC energy, implying that 
within our JAGP wave function it is possible  to improve substantially  
the nodes of the  HF Slater determinant, with a little extension 
of the variational freedom. 
It is important to emphasize that we have also 
optimized the HF determinant in the presence of the  
Jastrow factors described in Sec.~\ref{wave_function}. 
On the other hand, as shown in Tab.~\ref{table_C}, 
we have obtained the HF energy  
within our general Monte Carlo optimization scheme, even though, in this 
case, it is obviously not necessary to use a statistical method.
The LRDMC calculation  in the HF  case was done after 
optimizing  the two-body and three-body Jastrow factors, without 
changing the HF determinant.
Although the variational energy of this Slater-Jastrow wave function  
is higher than the corresponding fully optimized HF+Jastrow one, their LRDMC  
energies are the same. This implies that,  within a single determinant 
wave function, it is difficult in this case to improve the nodes even when 
the Jastrow and the determinantal parts are optimized together.

Nevertheless, our optimization scheme 
is very stable and reliable and allows to optimize a rather large number of 
variational parameters in a systematic way.
Within the JAGP ansatz   and in particular for the benzene molecule, 
it is extremely important to optimize the wave function in order 
to improve the nodal structure, and obtain a good LRDMC total energy. 
This was previously pointed out by two of us, 
in an all-electron calculation within the standard DMC
framework.\cite{casulamol}  
In that case, the DMC method
provides the same total energies as the LRDMC method, used here, since for  
both methods  
the FN approximation is  exactly the same in absence of pseudopotentials.
 
\begin{table*}
\caption {\label{table_C} LRDMC ground state energies (Hartree units) 
 for carbon pseudoatom  
using various basis sets. The LRDMC value is a rigorous upper bound 
for the ground state energy. The limit $a\to 0$ was obtained by using
$\eta=1+3.2 a^2$ in Eq.~\ref{laplace}.
The 2-body Jastrow factor has the form reported in Eq.~\ref{two-body}. 
The Jastrow and the AGP (or HF) geminals are 
expanded on a primitive gaussian basis denoted by $(ns~mp)$, where $n$ 
($m$) is  the number of s-wave (p-wave) gaussian orbitals.
Analogously, the number and type of contracted orbitals follow 
the slash symbol.
In particular $(6s6p)/[1s1p]$  denotes the standard HF Slater 
determinant. For comparison the HF energy  obtained 
in the CBS limit is  $-5.319505$ Hartree. 
}
\begin{ruledtabular}
\begin{tabular}{|c|c|c|c|c|}
 Wave function   & 3-body J basis & AGP basis   &   VMC  &   
  LRDMC      \\
\hline
\hline
AGP+2-body & -   &  (2s2p) &   -5.266 (1)  & -5.397(1)  \\
\hline
AGP+2-body & -  & (3s3p) &     -5.392 (1)  & -5.416(1)  \\
\hline
AGP+2-body & - & (4s4p) & -5.4066(4)  &  -5.4178(3)  \\
\hline
AGP+2-body & - & (5s5p)  & -5.4095(3)  &  -5.4180(1) \\
\hline
AGP+2-body & - & (6s6p)  & -5.4096(2)  &  -5.4181(1) \\
\hline
AGP+2-body & - & (5s5p1d)   & -5.4096(2) &  -5.4182(1) \\
\hline
AGP+2\&3-body & (1s1p) & (5s5p) & -5.4103(2) & -5.4181(1) \\ 
\hline
HF & -  & (6s6p)/[1s1p]  & -5.3193(3) &  -5.4107(3)  \\
\hline
HF+2\&3-body & (1s1p)  & (6s6p)/[1s1p]  & -5.3991(3) &  -5.4107(2)  \\
\hline
AGP+2\&3-body & (2s)  & (6s6p)/[2s1p]  &  -5.4075(2) &   -5.4160(1)  \\
\hline 
AGP+2\&3-body & (3s2p) & (4s5p)/[2s2p] &  -5.4115(1) &   -5.4182(1) \\ 
\hline 
AGP+2\&3-body & (3s2p) & (6s6p) &  -5.4113(1) &   -5.4183(1) \\ 
\end{tabular}
\end{ruledtabular}
\end{table*}

\begin{table*}
\caption {\label{table_O}   
Same as in Tab.(\ref{table_C}) for the oxygen pseudoatom. For a comparison 
with the reported values, the unrestricted HF, the MP2 and the CCSD(T) 
on the VTZ basis set have total energies of $-15.7149$, $-15.8636$, and $-15.8822$ 
respectively, calculated with Gaussian 03, Revision C.02~\cite{gaussian}. 
The limit $a\to 0$ was obtained by using
$\eta=1+10.8 a^2$ in Eq.~\ref{laplace}.}
\begin{ruledtabular}
\begin{tabular}{|c|c|c|c|c|}
 Wave function   & 3-body J basis & AGP basis   &   VMC  &   
  LRDMC      \\
\hline
\hline
AGP+2-body & -   &  (2s2p) &   -15.410 (3)  & -15.834(2)  \\
\hline
AGP+2-body & -  & (3s3p) &     -15.813(1)   &  -15.884(1)  \\
\hline
AGP+2-body & - & (4s4p) & -15.8611(6)  & -15.8901(3)  \\
\hline
AGP+2-body & - & (5s5p)  & -15.8687(4)  &  -15.8916(2) \\
\hline
AGP+2-body & - & (6s6p)  & -15.8685(6)   &  -15.8918(3) \\
\hline
AGP+2-body & - & (5s5p1d)   & -15.8679(5)  &  -15.8920(3) \\
\hline
HF+2-body & - & (5s5p)/[1s1p]   & -15.8674(5)  &  -15.8920(3) \\
\end{tabular}
\end{ruledtabular}
\end{table*}

\subsection{Binding energy of $C_2$ and benzene molecule, 
a size consistency study}
\label{size_consistency}

Though the $[2s1p]$ contraction is a rather small basis and does not provide the 
converged result in the total energy of the carbon atom, it represents 
a good compromise between accuracy and efficiency, because it can 
describe satisfactorily the chemical bond in all carbon-based 
compounds studied, as it is shown in Tab.~\ref{binding}.

To this purpose, in this Table we have reported two methods 
to calculate the binding energy. 
In the standard method (method I), we compute the difference between the 
total energy at the equilibrium distance and the 
sum of the energies of the independent fragments for a chosen atomic basis set.
The second method (method II) is based on the evaluation of 
the difference between the total energy at the equilibrium distance and the 
energy directly obtained when the constituents of the compound are still together, but pulled apart at large distance. 

In the following we show that method II is more appropriate
to compute the binding energy within the JAGP ansatz.
Indeed, the AGP part is the particle conserving BCS version 
only for the total number, not for the number in a local sector of the wave function.  
Therefore, if more than one fragment is included in the same AGP wavefunction,
the number of electrons  on each fragment is 
not conserved, and this leads to unphysical charge fluctuations which are energetically expensive. The Jastrow factor can significantly lower the energy, by imposing the right occupation number, but the local conservation of charge is fully restored only in the CBS limit of the Jastrow expansion. Thus, 
with a finite basis set in the Jastrow factor, the JAGP wavefunction
is clearly more accurate for a single fragment 
than for the whole system,  and method I usually underestimates the binding energy of the compound.
On the other hand, method II is much more accurate, as it includes the cancellation of 
the finite basis set errors in the Jastrow term.

Moreover, in order to exploit a better cancellation of errors, 
it is important that the energy of the  
fragments at large distance is obtained by 
iteratively optimizing the wavefunction of the compound  for 
larger and larger 
separations of the fragments. In this way, one follows adiabatically the 
fragmentation process, and avoids possible 
spurious energy minima, that may  occur in the optimization of a non linear 
function such as the JAGP.

For a perfectly size consistent wave function, methods I and II 
should coincide in the CBS limit. 
The JAGP wavefunction is perfectly size consistent  
for fragments which are singlet, and with the Jastrow factor
in the CBS limit. In the case of two singlets at large distance, it is 
enough to define the matrix $\lambda$ of the compound as the
sum of the two fragments $A$ and $B$  ($ \lambda= \lambda^{A,A}+ \lambda^{B,B}$),
with an appropriate Jastrow factor freezing the charge in $A$ 
and and $B$ when these two are far apart. 
In presence of unpaired orbitals, e.g. for the triplet Carbon atom, 
size consistency is very difficult to fulfill in general.
For instance,  a singlet $S=0$ $C_2$ wavefunction 
corresponding to two entangled Carbon atoms at large distance
can be obtained only with six independent Slater determinants, by appropriately combining the  two unpaired $p-$orbitals of each Carbon HF wavefunction, 
i.e.:
\begin{eqnarray}
&& | S=0,  {\rm A~far~from~B} \rangle =
\nonumber \\
&& \frac{1}{\sqrt{3}} ~
| p_x \uparrow A, p_y \uparrow A,  p_x \downarrow B, p_y \downarrow B \rangle  \nonumber \\ 
&+&  
\frac{1}{\sqrt{3}} ~
| p_x \downarrow A, p_y \downarrow A , p_x \uparrow B, 
p_y \uparrow  B \rangle  \nonumber \\
 &-& \frac{1}{2\sqrt{3}} ~
 | p_x \uparrow A, p_y \downarrow A , p_x \uparrow B, p_y \downarrow B \rangle \nonumber \\
&-& \frac{1}{2\sqrt{3}} ~ 
| p_x \uparrow A, p_y \downarrow A , p_x \downarrow B, p_y \uparrow B \rangle   \nonumber \\
&-& \frac{1}{2\sqrt{3}} ~
| p_x \downarrow A, p_y \uparrow A , p_x \uparrow B, p_y \downarrow B \rangle \nonumber \\
&- & \frac{1}{2\sqrt{3}} ~
| p_x \downarrow A, p_y \uparrow A , p_x \downarrow B, p_y \uparrow B \rangle  \label{expc2}
\end{eqnarray}  
where each term in the above expression is a single determinant,
with  the orbitals indicated inside the brackets, toghether with 
the four $2s$ orbitals ($2s  \uparrow  A $, $2s  \downarrow  A $, 
 $2s \uparrow  B$, $2s  \downarrow  B $).

The JAGP wavefunction can be perfectly 
size consistent even for triplet fragments in the ideal but important limit 
of strong repulsion between electrons in the same orbital (strong Hubbard $U$). 
In this limit the occupation of  the same unpaired $p-$orbital by electrons of opposite 
spins is forbidden as in the singlet expansion for $C_2$ (Eq.~\ref{expc2}), and the two Carbon Slater 
determinants, each with two unpaired orbitals $p_x$ and $p_y$,  can be joined in a single determinant 
(AGP singlet), by turning on matrix elements such as: 
$$\lambda^{p_x,p_x}_{A,B} = \lambda^{p_y,p_y}_{A,B}= 
\lambda_{A,B}^{p_x,p_y} =- \lambda^{p_y,p_x}_{A,B},  $$
where $A$ and $B$ indicate the two Carbon atoms at large distance, and 
these matrix elements are assumed to be small compared with  the occupied 
$2s$ orbitals (e.g. $\lambda^{2s,2s}_{A,A}=\lambda^{2s,2s}_{B,B}=1$).
Then, it is simple to show that the $6$ Slater determinants defining the $C_2$ 
singlet can be obtained with the correct coefficients 
by { \em a single determinant}  AGP wavefunction,
provided the double occupations of the $p_x$ and $p_y$ orbitals can be 
 projected out by an appropriate  Jastrow factor. 
However,  the Jastrow factor used here, 
within the present JAGP expansion, can only partially project out double 
occupation of the same orbital because it  depends only on  the 
total electron density and not  
explicitly on the corresponding 
  angular momentum orbital components. 
By consequence,  the present JAGP wavefunction 
can be only approximately size consistent in this special case. 
However, since this  loss of size consistency is clearly due to a local 
effect of the correlation on the same atomic orbital,
one expects that this contribution should 
be almost the same both at the equilibrium and at large $A-B$ distance, 
and therefore it should affect weakly the chemical bond.
 
\begin{figure}[!ht]
\includegraphics[width=\columnwidth]{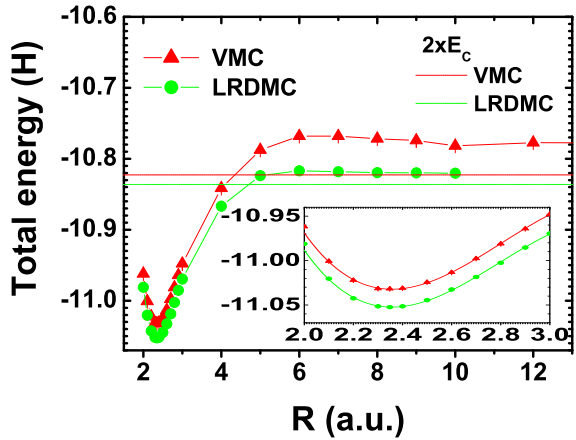}
\caption{\label{evsrc2} Energy for two Carbon atoms 
as a function of their distance. We used a (4s5p)/[2s2p] GTO basis set for the AGP part, and a (3s2p) uncontracted GTO basis set for the Jastrow factor. The atomic basis set convergence has been reached within $1mH$ at the VMC level for the molecule at the 
equilibrium distance.
The LRDMC and VMC energies are not fully size consistent 
(for large $R$ they should approach the energies  indicated by the full lines standing below). 
The VMC curve shows a maximum at $R=6$ and a shallow minimum at 
$R=10$,  which are almost completely 
removed by the LRDMC energies (within an accuracy of $0.1eV$). 
The inset shows an expansion of the picture around the equilibrium distance. 
Here a cubic polynomial has been used for  fitting the data 
in the range $ 2.1 \le R \le 3$. From this interpolation 
the resulting VMC [LRDMC] equilibrium distance is 
$2.357(4)~a.u.$ [$2.358(5)~a.u.$], and the ZPE is 
$4.20(4)~mH$ [$4.20(5)~mH$].
}
\end{figure}

Within this hypothesis, that looks very well confirmed in 
Fig(\ref{evsrc2}), it is possible to obtain a good chemical accuracy 
($\simeq 0.1eV$), by using only a single geminal JAGP ansatz. 
Notice that in this picture the LRDMC energy appears very smooth and 
reasonable at large distance, even without approaching the energy of 
two isolated Carbon atoms. This calculation suggests that in the
exact size consistent framework, which includes  
many AGP or determinants, the total 
energy should acquire only an irrelevant rigid shift, at least 
within the LRDMC method.
Unfortunately we are not aware of very 
accurate calculation of the full energy 
dispersion in $C_2$, but the zero point energy (ZPE), computed 
from the data in Fig.(\ref{evsrc2}), is in very good agreement 
with the experimental value ($4.2mH$)\cite{nonso}, clearly supporting the 
accuracy of our calculation apart for an irrelevant energy shift.  

The very remarkable outcome of this careful analysis is that 
it is possible to describe well the chemical bond in most of the
interesting carbon-based compounds, as shown in Tab.~\ref{binding}. 
As a further confirmation that this hypothesis is plausible,
we have computed the  equilibrium distance of the carbon dimer (Tab.~\ref{Rminc2}) using the simultaneous optimization of the bond length and the variational parameters as 
described in Ref.~\onlinecite{casulamol}. 
By means of this technique, based on energy derivatives, we can compute the 
bond length much more accurately than by fitting only the total energy around the minimum (e.g. in the calculation of the bond length in Fig.~\ref{evsrc2}, 
a much larger statistical error is obtained for this quantity). We found a perfect agreement with the experimental bond length in the large basis set limit. 
\begin{table*}
\caption {\label{binding} 
Binding energy (eV) for carbon-based compounds, obtained with the JAGP 
wave function described in the text for a given atomic basis set, reported in
the table. The most accurate binding energy is 
obtained by evaluating the difference between the total energy at the equilibrium 
and at large distance (method  II). For the benzene molecule 
we exploited the size consistency of the JAGP ansatz valid for singlet  
fragments and complete Jastrow factor. Therefore, 
we considered first the fragmentation process $ C_6 H_6 \to 3 H_2 + 3 C_2$, 
and then we used the already determined $C_2$ binding energy with method 
$II$ (for $H_2$ the JAGP is clearly size consistent, since it is exact for 
two electrons).
The less accurate method is the standard one (method I), obtained   
by computing the large distance energy 
by summing  the energy  of each individual fragment with the same basis.  
In the contracted  $[2s2p1d^*]$, $[2s2p^*]$  or   $[2s1p^*]$ cases,
 used generally here  in the large system cases, 
the coefficients of the contracted orbitals 
are   assumed to be independent of the angular momentum projection $l_z$.  
Notice also that the inclusion of the polarization d-orbital 
does not affect the binding of $C_2$  within $1mH$.
The DMC HF binding energy (I) for $C_2$ is 
$5.66eV$\protect\cite{cyrc2}. The last column refers to the non relativistic 
value estimated either by experiments or by a very accurate calculation for 
$C_2$.}
\begin{ruledtabular}
\begin{tabular}{|c|c|c|c|c|c|c|c|c|}
 Compound    &  3-body J basis & AGP basis &   \#  par &
  VMC (I) &  LRDMC (I) & VMC(II)  & LRDMC (II)   & Estimated  \\
\hline
$C_2$ &  (3s2p) & (6s6p)/[2s1p$^*$] &  69 & 5.806 (16)  & 5.946(4) & 6.766(25)  &  6.267(4)  &  6.36(1) \footnotemark[1] \\ 
\hline
$C_2$ &  (3s2p) & (6s6p)/[2s1p]  &  74 & 5.884(16) & 5.959(4) & 6.862(6) &  6.283(5)  &  6.36(1) \footnotemark[1] \\ 
\hline 
$C_2$ &  (3s2p) & (4s5p)/[2s2p$^*$] & 95 & 5.688(8) & 5.883(4) & 6.910(7) & 6.318(9) & 6.36(1) \footnotemark[1] \\ 
\hline 
$C_2$ &  (3s2p1d) & (4s5p1d)/[2s2p1d$^*$] & 136  & 5.724(8) & 5.887(4) & 6.893(7) & 6.314(7) & 6.36(1) \footnotemark[1] \\ 
\hline
$C_2$ &  (3s2p) & (6s6p) &  255 & 5.763(12) & 5.812(4) &  6.737(5) &  6.289(4) &   6.36(1) \footnotemark[1] \\ 
\hline
\hline
$C_6 H_6$ & (3s2p) & (6s6p)/[2s1p$^*$] & 505 & 57.06(3) &  58.105(9) &  59.942(60)  &  59.067(8) &   59.24(11) \footnotemark[2] \\ 
\end{tabular}
\footnotetext[1]{Ref.~\onlinecite{cyrc2}}
\footnotetext[2]{Ref.~\onlinecite{ermler}}
\end{ruledtabular}
\end{table*}

\begin{table}
\caption {\label{Rminc2} 
Equilibrium distance of the $C_2$ molecule obtained by minimizing 
the energy of the JAGP with the given basis  set.  The symbols used refer 
to the ones defined in previous tables.}
\begin{ruledtabular}
\begin{tabular}{|c|c|c|c|c|}
3-body J basis & AGP basis & \#  par &  R (VMC)  & R (exp)    \\
\hline
\hline
 (3s2p) & (6s6p)/[2s1p$^*$] &  69 & 2.3555(8)    &  2.3481  \\ 
\hline
(3s2p) & (6s6p)/[2s1p]  &  74 & 2.3559(9)  & 2.3481  \\ 
\hline
(3s2p) & (6s6p) &  255 & 2.3480(6)  & 2.3481  \\ 
\end{tabular}
\end{ruledtabular}
\end{table}

\subsection{The benzene dimer}
\label{benzene_dimer}
As discussed in the previous subsection for two singlet molecules A and B  with 
electron number $N_A$ and $N_B$ respectively, the JAGP is size consistent 
whenever the three-body Jastrow factor is optimized in the CBS limit.
In this way, this term can fully project out the charge fluctuations present in the AGP 
part of the wave function, which would erroneously allow  a number of 
electrons different from $N_A$ and $N_B$ even when the molecules A and B are 
at very large distance.  
In our variational wave function the Jastrow geminal (Eq.~\ref{3body}) 
is defined only on a (3s2p) single zeta 
gaussian  basis set for the carbon atom, and a (1s) single zeta for the
hydrogen atom.
Nevertheless, the wave function is very close to be 
size consistent, because the total 
energy evaluated at a fairly large distance, i.e. $12~a.u.$, 
is given by $E_{A+B}=-75.0825 \pm 0.0003 H$ after a full 
energy optimization,  
whereas the energy of a single benzene molecule within the same basis set
is given by  $E_{A}=-37.5422 \pm 0.0002 H$, i.e. exactly $E_{A+B}/2$ 
within error bars. Therefore, the JAGP ansatz with the chosen basis set 
is supposed to be accurate enough to describe the weak interactions in the benzene dimer, 
as both the basis set convergence and the size consistent behavior are taken
into account.

\begin{figure}[!ht]
\includegraphics[width=\columnwidth]{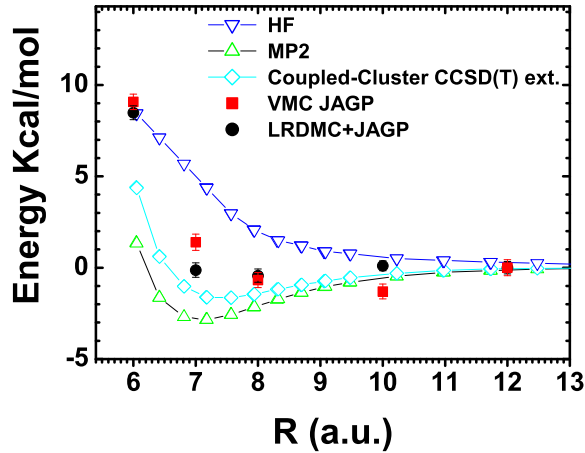}
\caption{\label{dimpar} Energy for two face-to-face benzene molecules 
as a function of their distance for different methods.  
The reference was taken at $R=12$. The LRDMC kinetic parameters are
$\eta=1.8$, $a=0.5 a.u.$, and $a^\prime/a=\sqrt{5}$.
The nearest neighbor $C-C$ ($C-H$) distance was set 
to $2.636$ ($2.038$) a.u. in the two molecules.}
\end{figure}

The full dispersion curve of the benzene dimer is reported in
Fig.(\ref{dimpar}) for a face-to-face 
geometry, together with the more accurate LRDMC results. 
As it is apparent from this picture, the 
LRDMC result does not change qualitatively  the variational outcome, showing 
a very weak dispersion, much less deep if compared to the 
most accurate CCSD(T) results. Our best value of the binding energy is
$0.5(3)$ kcal/mol. 
It is possible that the LRDMC method 
reduces the VMC bond length  ($9-10 a.u.$) 
by  $1-2 a.u.$, though an accurate determination of this quantity is 
rather difficult due to the very shallow minimum.

\begin{figure}[!ht]
\includegraphics[width=\columnwidth]{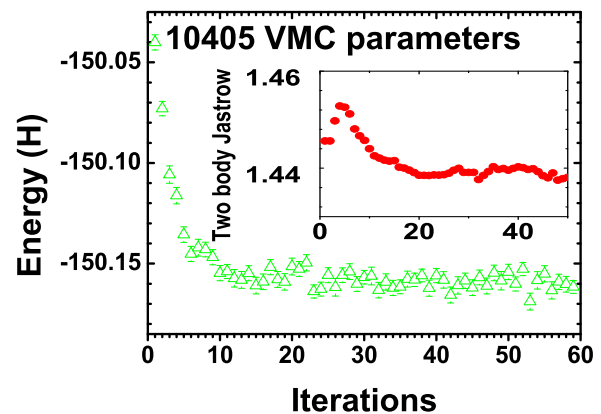}
\caption{\label{minwr} 
Total energy (Hartree) of the variational wave function during the 
optimization of all the $10405$ variational parameters 
consistent with the chosen basis
in the parallel displaced geometry shown in Fig.(\ref{c12h12pd}).
The case $R_1=7~a.u.$ and $R_2=3.4~a.u.$ is considered here. In the inset the 
evolution of the variational parameter $F$ (Eq.~\ref{two-body}) is shown.
}
\end{figure}

\begin{figure}[!ht]
\includegraphics[width=\columnwidth]{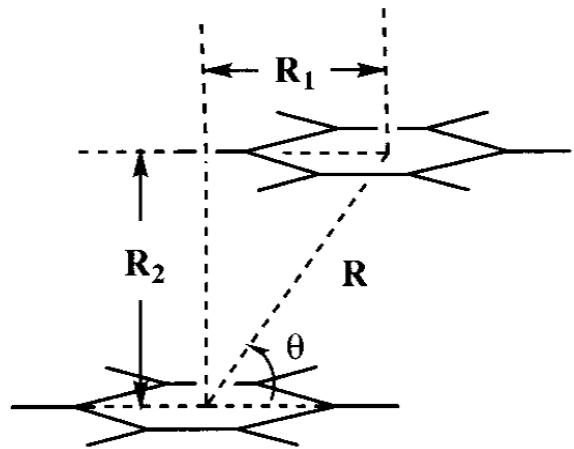}
\caption{ 
 Geometry of the benzene dimer with the $R_1$ and $R_2$ distances studied in
 this work.\label{c12h12pd}}
\end{figure}

We have extended the calculation to the parallel displaced geometry 
(see Fig.\ref{c12h12pd}), which has been proposed 
 to be the most stable configuration.
However, since in this case the number of variational parameters is larger
($\simeq 10000$), we have used partial information of the Hessian matrix, 
following the scheme introduced in Ref.~(\onlinecite{srh}) to accelerate
the convergence of the minimization.
In the first iteration we move the parameters along
the direction $\vec g_1=s^{-1}\vec  f$, with $s$ the regularized
SR matrix in Eq.~\ref{regularization}. At this step, 
the Hessian matrix gives the optimal amplitude $\gamma_1$ of  the
parameter change $\delta \vec \alpha = \gamma_1 \vec  g_1 $.
Analogously, after $n$ iterations 
the variation of  the  parameters is given by 
$\delta \vec \alpha = \sum_{i=1}^n \gamma_i \vec g_i $, where 
$\{\gamma_i\}_{i=1,n}$ are  determined by using the Hessian matrix 
of the last iteration $n$. After 
changing the variational parameters $ \vec \alpha_k \to \vec \alpha_k+ 
\delta\vec  \alpha_k$, a new vector $\vec g_{n+1}=s^{-1}\vec  f $ 
is computed with the new wave function, and then the procedure is repeated
iteratively. Notice that a single optimal direction is ``collective'' in the
parameter space, as it involves many degrees of  freedom. 
In this way the minimization proceeds in a very stable and fast way,
as shown in Fig.~\ref{minwr}. The main advantage of this method  is that
the Hessian matrix can be calculated in a small basis 
set, and it is not necessary to introduce 
any further regularization parameter other than $\epsilon=10^{-4}$ (Eq.~\ref{regularization}).

The optimization of the parallel displaced benzene dimer is rather heavy
(about two  days on a 64 processor SP5 parallel machine) 
because in every iteration shown in Fig.(\ref{minwr}) a 
very high statistical accuracy is required, 
due to the so many variational parameters, 
otherwise all the matrices involved in the iteration process (especially the 
large overlap matrix $s$) are too much noisy. 
For this reason we have performed the wave function optimization 
only for two  particular geometries reported in Tab.~\ref{bindingbe}.
From  the force components in the two inequivalent directions, 
it is clear that the minimum energy  occurs at a value of
$R_2=7.5  \pm 0.2 a.u.$, while $R_1$ is about unchanged. The binding energy is 
$2.2(3)$ kcal/mol.

\section{Conclusion}
In  this work, we have devised a QMC framework which is able to provide
reliable estimates of weak chemical bonds, mainly driven by vdW dispersive
forces. We used a  
highly correlated variational wave function, the JAGP ansatz, which contains
all the necessary ingredients to describe intermolecular interactions: (i) a very
high ``on site'' accuracy, through the inclusion of near degeneracy
correlation effects in the AGP part, 
(ii)  the possibility to control the molecular charge
distribution  through a local 3-body Jastrow factor, (iii)  the capability to
take into account the intermolecular correlation, responsible for the weak
dispersive forces, by means of a ``long-range'' Jastrow  term,  which connects
the molecules (or the fragments) involved in the interaction.
Although the JAGP ansatz is not size consistent in general,
we have shown that in the carbon-based compounds analyzed here it is possible
to obtain accurate and reliable  results, by taking 
the calculation of the system in the large distance geometry as reference point.

We have described an improved optimization method, based on a proper
regularization of the overlap matrix in the SR scheme,
which can be further boosted by the information of the Hessian matrix.
With this method it is possible to optimize a number of parameters of the
order of $10000$. Our fully optimized variational wave function has been used 
as initial guess in projection LRDMC calculations. We also found the optimal
setting of the kinetic parameters in the LRDMC method, in order to speed up
the diffusion MC simulations with pseudopotentials.
After the optimization step and the LRDMC projection, our results are very weakly
dependent on  the basis set used, at variance with the post-HF quantum chemistry
methods.

We studied the face-to-face and displaced parallel geometry and energetics 
of the benzene dimer, which is a prototype compound to understand
intermolecular dispersive 
forces. After a full optimization of both the Jastrow and the 
AGP part, the VMC binding energy remains in qualitative agreement 
with the LRDMC result, which is supposed to be the most accurate
QMC calculation. All these findings strongly support the reliability of our
numerical study. 

The binding of the benzene dimer appears small 
and almost negligible ($\simeq 0.5$ kcal/mol) in the face-to-face geometry.  
On the other hand, in the  parallel displaced configuration
where the two molecules are shifted by a distance  $R_1=3.4 a.u.$, 
there is a sizable gain in energy, 
which reaches its optimal value of $~2.2(3)$ kcal/mol at $R_2 \simeq 7 a.u.$.
Apparently, this is smaller than the most recent post HF value ($~2.8$
kcal/mol \cite{cc}) obtained with the $CCSD(T)$ method, after
a careful extrapolation to the CBS limit. However, 
by considering the reduction of the binding energy due to the 
zero point vibrational energy $ZPE$ ($\Delta ZPE=0.37$kcal/mol), 
our result goes clearly in the direction of the best experimental estimate 
of the binding energy, which is $1.6 \pm 0.2$ kcal/mol.\cite{bestexp}. The
agreement between the experiment and our theoretical prediction is another
striking sign of the capability of the QMC techniques to describe accurately
not only a strong intramolecular bond, but also 
the very weak intermolecular attractions based on vdW dispersive forces.

\begin{table}
\caption {\label{bindingbe} 
Binding energies $\Delta E$ (kcal/mol) 
and forces (kcal/(mol a.u.)) acting on the two independent directions
$\vec R_1$ and $\vec R_2$ shown in Fig.~\ref{c12h12pd}. Energies differences
are evaluated  with respect to the large separation geometry
($R_1=0$,$R_2=12  a.u.$), used also in Fig.~\ref{dimpar}. The 
forces are computed in a VMC calculation with the optimized variational
wave function, and include both Feynman and Pulay contributions.
 $R_1$ and $R_2$ are given  in a.u. }
\begin{ruledtabular}
\begin{tabular}{|c|c|c|c|c|c|}
 $R_1$ & $R_2$  &  $F_1$ &  $F_2$ & $\Delta E_{VMC}$  &$\Delta  E_{LRDMC}$ \\
 \hline  
  0  &   7     &    0        &   2.1(2)    &   -1.4(4)  &  0.2(3) \\   
\hline
  0  &   8     &    0        &   0.1(2)   &     0.7(3)  &   0.5(3)  \\
  \hline 
 3.4 &   7     &   0.20(8) & 0.6(1)  &     1.4(3)  &   2.2(3)  \\ 
 \hline  
 3.4 &   8     &   -0.22(6) & -0.7(1) &    2.0 (3)  &  1.8(3)   \\  
\end{tabular}
\end{ruledtabular}
\end{table}

\acknowledgments 
We thank Leonardo Guidoni and Giacinto Scoles for helpful discussions, and Claudia Filippi for sending 
us accurate pseudopotentials for oxygen and carbon atoms.
We are also indebted to J. Toulouse and C. Umrigar for their careful reading of
this manuscript. 
This work was partially supported by COFIN2005, and CNR. 
One of us (M.C.) acknowledges support in the form of the NSF grant DMR-0404853.


\begin{thebibliography}{99}
\bibitem{hopping} With probability $1/2$ we perform the standard Metropolis 
algorithm: in this case each electron coordinate 
is moved  only in a neighborhood of the initial position $\vec r_i$.
With the same probability  we make a large move in the direction of an
arbitrary atom $R_j$ ( chosen  
randomly), by a vector equal to $\vec R_j-\vec R_{min}$ 
 where $\vec R_{min}$ represents  the  
atomic  position closest to $\vec r_i$.
\bibitem{note3} This diagonal matrix is positive definite because 
every diagonal element  $s_{i,i}> 0$, as  $s$ is positive definite.
\bibitem{notelrdmc} In practice we set 
the locality parameter $\alpha=0$ in the LRDMC scheme\cite{lrdmc}.
\end{thebibliography}
\end{document}